\documentclass[%
preprint,
 amsmath,amssymb,
 aps,
]{revtex4-1}

\usepackage{graphicx}
\usepackage{dcolumn}
\usepackage{bm}

\usepackage{amssymb}
\usepackage{amsmath}
\usepackage{amsfonts}
\usepackage{color}

\newcommand{\benu}{\begin{enumerate}}
\newcommand{\eenu}{\end{enumerate}}
\newcommand{\beq}{\begin{equation}}
\newcommand{\eeq}{\end{equation}}
\newcommand{\beqn}{\begin{eqnarray}}
\newcommand{\eeqn}{\end{eqnarray}}
\newcommand{\beqd}{\begin{eqnarray*}}
\newcommand{\eeqd}{\end{eqnarray*}}
\newcommand{\bea}{\begin{array}}
\newcommand{\eea}{\end{array}}
\newcommand{\bcen}{\begin{center}}
\newcommand{\ecen}{\end{center}}
\newcommand{\btab}{\begin{tabular}}
\newcommand{\etab}{\end{tabular}}
\newcommand{\bsub}{\begin{subequations}}
\newcommand{\esub}{\end{subequations}}

\newcommand{\beit}{\begin{itemize}}
\newcommand{\enit}{\end{itemize}}

\begin{document}


\title{Influence of pairing correlations on the radius of neutron-rich nuclei}

\author{Ying Zhang}
\email{yzhangjcnp@tju.edu.cn}
\affiliation{Department of Physics, School of Science, Tianjin University, Tianjin 300072, China} 
\author{Ying Chen}
\affiliation{Institute of Materials, China Academy of
Engineering Physics, Mianyang 621700, China}
\author{Jie Meng}
\affiliation{State Key Laboratory of Nuclear Physics and Technology, School of Physics, Peking University, Beijing 100871, China}
\affiliation{School of Physics and Nuclear Energy Engineering, Beihang University, Beijing 100191, China}
\affiliation{Department of Physics, University of Stellenbosch, Stellenbosch, South Africa}
\author{Peter Ring}
\affiliation{State Key Laboratory of Nuclear Physics and Technology, School of Physics, Peking University, Beijing 100871, China}
\affiliation{Fakult\"at f\"ur Physik, Technische Universit\"at M\"unchen, D-85748 Garching, Germany}

\date{\today}

\begin{abstract}
The influence of pairing correlations on the neutron root mean square (rms)
radius of nuclei is investigated in the framework of self-consistent Skyrme
Hartree-Fock-Bogoliubov calculations. The continuum is treated appropriately
by the Green's function techniques. As an example the nucleus $^{124}$Zr is
treated for a varying strength of pairing correlations. We find that, as the
pairing strength increases, the neutron rms radius first shrinks, reaches a
minimum and beyond this point it expands again. The shrinkage is due to the
the so-called `pairing anti-halo effect', i. e. due to the decreasing of the
asymptotic density distribution with increasing pairing. However, in some
cases, increasing pairing correlations can also lead to an expansion of the
nucleus due to a growing occupation of so-called `halo' orbits, i.e. weakly
bound states and resonances in the continuum with low-$\ell $ values. In
this case, the neutron radii are extended just by the influence
of pairing correlations, since these `halo' orbits
cannot be occupied without pairing. The term `anti-halo effect' is not
justified in such cases. For a full understanding of this complicated
interplay self-consistent calculations are necessary.

\end{abstract}

\maketitle


\section{\label{sec:intro}Introduction}
Superfluidity is a quantum phenomenon found in various systems such as
liquid helium, superconductors, atomic nuclei, and neutron stars. Nuclear
superfluidity is caused by pairing correlations, induced by the attractive
effective interaction between pairs of nucleons. This leads to an odd-even
staggering in nuclear masses and separation energies, and to a considerable
reduction of the moments of inertia in rotational bands. These phenomena are
observed throughout the entire periodic table~\cite{Brink2005}. In the past
two decades, exotic nuclei with large proton or neutron excess have been
extensively discussed and new phenomena have been discovered such as proton
radioactivity close to the proton drip line or neutron halos in some nuclei
at the neutron drip line. The coupling to the continuum plays an essential
role in these weekly bound
systems~\cite{Bertsch1991_APNY209-327,Hansen1987_EPL4-409,Meng1996_PRL77-3963,Meng1998_PRL80-460,Meng1998_NPA635-3}. A famous case is the nucleus $^{11}$Li, where the first neutron halo has
been observed~\cite{Tanihata1985_PRL55-2676}. Without pairing correlations
the two neutrons in the halo would not be bound to the $^{9}$Li core.
In these nuclei close to the neutron drip line the Fermi energy approaches the continuum threshold and pairing correlations make it possible for neutrons to occupy not only the weakly bound orbits but also unbound orbits with very low orbital angular momentum $\ell =0$ or $\ell =1$ near the Fermi energy in the single-particle spectrum~\cite{Lalazissis1998_NPA632-363}. This could be most easily seen in the canonical basis, where the HFB wave function can be represented in the form of a BCS-state as it is discussed in detail in Refs.
~\cite{Meng1996_PRL77-3963, Meng1998_PRL80-460, Meng1998_NPA635-3,CHEN-Y2014_PRC89-014312}.
Their wave functions
can extend far outside the nucleus due to the low centrifugal barrier, which
is crucial to the formation of the halo structure~\cite{Meng1996_PRL77-3963}.

Of course, there is also the possibility to form a halo without pairing. If,
for instance, the last occupied neutron orbit has an orbital angular
momentum $\ell=0$ and a single-particle energy $\varepsilon$ just below the
continuum threshold, in the asymptotic region the dominant contribution to
the Hartree-Fock (HF) density of this nucleus has the form
\begin{equation}
\rho_{\mathrm{HF}}(r) \propto \frac{\exp(-2\kappa r)}{ r^2}\quad\mathrm{for}%
\quad r\rightarrow\infty
\end{equation}
with $\kappa=\sqrt{2m|\varepsilon|}/\hbar$. The mean square neutron radius
calculated with this density behaves as
\begin{equation}
\langle r^{2}\rangle_{\mathrm{HF}} \propto\frac{1}{|\varepsilon|}.
\label{eq:rmsHF}
\end{equation}
It diverges for $\varepsilon\rightarrow0$.

Bennaceur \textit{et al.}~\cite{Bennaceur2000_PLB496-154} showed that
pairing correlations, leading to a finite pairing gap $\Delta ,$ prevents such
a divergence. The mean square radius calculated with the asymptotic
Hartree-Fock-Bogoliubov (HFB) density behaves as
\begin{equation}
\langle r^{2}\rangle _{\mathrm{HFB}}\propto \frac{1}{\Delta }.
\label{eq:rmsHFB}
\end{equation}%
Therefore, they concluded that the additional pairing binding energy acts
against a development of an infinite root mean square (rms) radius that
characterizes $\ell =0$ mean-field eigenfunctions in the limit of vanishing
binding energy. This is then called `pairing anti-halo effect'.

If one defines a `halo' by a divergence in the rms radius, this conclusion
is definitely mathematically correct. However in nature we have to consider
additional points: first, experimentally observed halos~\cite%
{Tanihata1985_PRL55-2676} have a large but finite rms radius; second, apart
from closed shells the coupling to the continuum causes and enhances pairing
correlations. Then the proper mean-field description of the nucleus is given
by the HFB theory.
In order to understand the structure of the HFB wave function, it would be useful to represent it as a BCS state in its canonical basis~\cite{Ring1980}, i.e. in terms of eigenstates of the density matrix, whose energies are defined as expectation values of the single-particle Hamiltonian $\hat{h}$ in the HFB equation. Then
one could find clearly that there are not only contributions to the radius from the canonical orbits below the continuum threshold of $\hat{h}$, but also from partially occupied states with energies in the continuum.
For orbits without or with small centrifugal
barriers, i.e. for $s$- or $p$-levels their contributions are not negligible~%
\cite{Meng1996_PRL77-3963}. The coupling to those states definitely grows
with increasing pairing correlations. If a halo could be formed only in
cases without pairing, i.e. without coupling to the continuum, the three
limiting conditions: zero pairing, $\varepsilon \rightarrow 0$, and low-$%
\ell $ values, would be actually difficult to meet in real nuclei due to the
shell structure.
Neutron halo phenomena would be very rare and rather accidental among the known neutron drip-line nuclei.

The influence of pairing correlations on single-particle configurations in
the continuum has been studied in the literature~\cite{Yamagami2005_PRC72-064308}. However, these
investigations were based on a fixed potential of Woods-Saxon shape and
the spatial extension of the density
was modified by changing artificially the depth of this
potential. The continuum was taken into account by solving  the
HFB
equation in a finite box of radius $R$.

It is the goal of this investigation to go a step further with respect to Ref.~\cite{Yamagami2005_PRC72-064308} and to clarify in a fully self-consistent way the influence of pairing correlations
on the extension of neutron radii in these nuclei close to the drip line,
as it has been observed in experiments (as for instance in Ref.~\cite{Tanihata1985_PRL55-2676}). It is certainly an interesting question to distinguish between the formation of a neutron skin or a neutron halo, as it has been done in several theoretical investigations~\cite{Mizutori2000_PRC61-044326, Rotival2009_PRC79-054308, Rotival2009_PRC79-054309}. However, it is not the goal of the present work to go into such details, as long as there are no precise experimental data available on the density distributions to distinguish these two phenomena.
In the
light of the above considerations it is evident, that the complicated
interplay of the different phenomena of changing mean fields and pairing
fields can only be achieved in fully self-consistent calculations with a
proper treatment of the continuum. We will consider not only the influence
of pairing on the asymptotic behavior of the wave functions of occupied,
weakly bound, low-$\ell $ orbits, but also the role of the occupation
probabilities introduced by the scattering of pairs around the Fermi energy
and the coupling to the continuum in these loosely bound superfluid systems.
Recently, covariant density functional theory has been used to study such
phenomena with a discretized continuum~\cite{CHEN-Y2014_PRC89-014312}.
Since, worldwide, non-relativistic density functional theory is one of the
most successful approaches in the description of exotic nuclei~\cite%
{Bender2003_RMP75-121} we concentrate here on investigations based on
density dependent HFB theory with Skyrme forces. A recently developed code
using the Green's function method~\cite%
{Belyaev1987_SJNP45-783,Matsuo2001_NPA696-371} allows us to avoid the
discretization of the continuum and to solve the continuum HFB equations
with the proper asymptotic behavior. Technical details of this method can be
found in Ref.~\cite{Zhang-Ying2011_PRC83-054301}.

\section{Technical details}
As an example we consider the nucleus $^{124}$Zr, which has been predicted to be a neutron halo nucleus
by relativistic continuum Hartree Bogoliubov
(RCHB) calculations~\cite{Meng1998_PRL80-460,ZHANG-SQ2003_SCG46-632}.
Similar results have been reproduced by the Skyrme HFB theory~\cite%
{Grasso2006_PRC74-064317,ZHANG-Ying2012_RPC86-054318}. We use the Skyrme
functional SkI4, which has been carefully adjusted to the isospin properties
of nuclear skins~\cite{Reinhard1995_NPA584-467}. It is therefore used in
many applications for the description of halo phenomena in the framework of
non-relativistic density functional theories, as in Refs.~\cite%
{Grasso2006_PRC74-064317,ZHANG-Ying2012_RPC86-054318,PEI-JC2006_NPA765-29}.
For the pairing force we use the density-dependent delta interaction (DDDI)
discussed in Refs. \cite{Matsuo2006_PRC73-044309,Matsuo2010_PRC82-024318}:
\begin{equation}
v_{q}^{\text{pair}}(\mbox{\boldmath$r$},\mbox{\boldmath$r$}^{\prime })=\frac{%
	1}{2}(1-P_{\sigma })V_{q}(\mbox{\boldmath$r$})\delta (\mbox{\boldmath$r$}-%
\mbox{\boldmath$r$}^{\prime }),~~~~(q=n,p),
\end{equation}%
where $V_{q}(\mbox{\boldmath$r$})$ is the pairing interaction strength. It
is a function of the neutron and proton densities as
\begin{equation}
V_{q}(\mbox{\boldmath$r$})=\eta V_{0}\left[ 1-x\left( \frac{\rho _{q}(%
	\mbox{\boldmath$r$})}{\rho _{c}}\right) ^{\alpha }\right] .
\label{eq:DDDI-vpeta}
\end{equation}%
In order to avoid the well-know ultra-violet divergencies, we work in a pairing
window, i.e. the quasiparticle space considered in these calculations is truncated
by the maximal orbital angular momentum $l=12\hbar $ and to the maximal quasiparticle
energy $E_{\text{cut}}=60$~MeV.
The parameters in Eq.~(\ref{eq:DDDI-vpeta}) are adopted as $V_{0}=-458.4$%
~MeV~fm$^{-3}$, $x=0.71$, $\alpha =0.59$, and $\rho _{c}=0.08$~fm$^{-3}$.
The parameter $V_{0}$ representing the strength of the
pairing force in free space is chosen to reproduce the scattering length $%
a=-18.5$~fm of the bare nucleon-nucleon interaction in the $^{1}S_0$ channel
\cite{Matsuo2010_PRC82-024318}. In order to study the influence on pairing
properties, we use in the following investigations the
additional factor $\eta $ which can change the pairing strength.

The contour integration path $C$ used for the calculation of the total densities
by the Green's function method of Ref. \cite{Zhang-Ying2011_PRC83-054301} is
chosen to be a rectangle with the height $\gamma =0.1$~MeV and the length $%
E_{\text{cut}}=60$~MeV. The energy step of the contour integration is $%
\Delta E=0.01$~MeV. For comparison, we also perform HFB calculations by the
box-discretized approximation, in which the HFB equation is solved with box
boundary conditions~\cite{Meng1996_PRL77-3963,Dobaczewski1996_PRC53-2809}.
Both the box-discretized and the continuum HFB calculation are performed
with a box size $R_{\mathrm{box}}=20$~fm, and a mesh size $\Delta r=0.1$~fm.

\section{Discussions}
\begin{figure}[ptb]
	\caption{Total energy $E_{\text{tot}}$ and neutron root mean square (rms)
		radius $r_{\mathrm{rms}}$ obtained by the continuum (filled circle) and the
		discretized (open circle) Hartree-Fock-Bogoliubov (HFB) calculations for $%
		^{124}$Zr with the SkI4 interaction as a function of the strength $\protect%
		\eta V_{0}$ of the DDDI pairing force, where $V_{0}=-458.4$~MeV~fm$^{-3}$.}
	\label{fig1}
	\centering
	\includegraphics[width=0.5\textwidth]{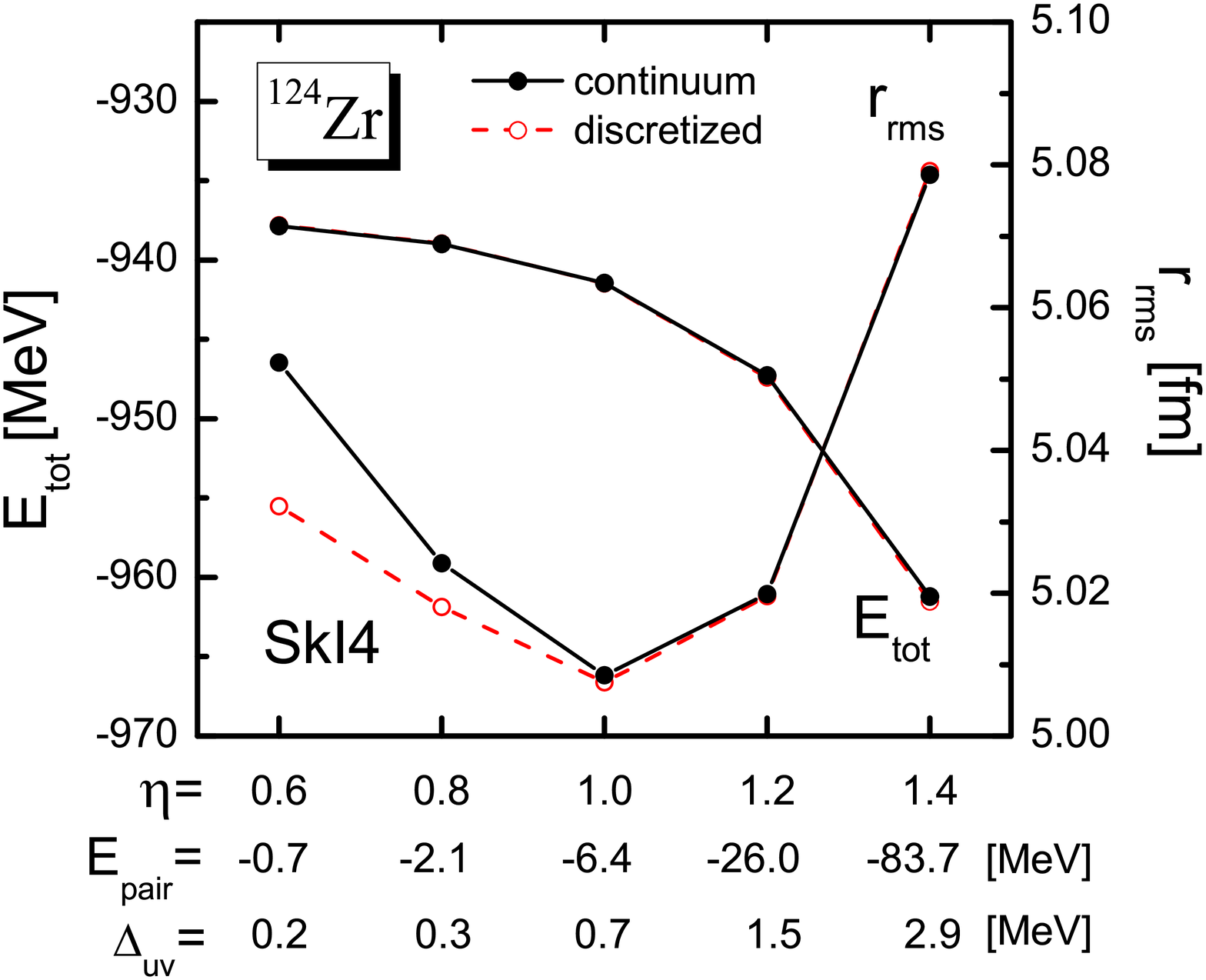}
\end{figure}

Fig.~\ref{fig1} shows the total energy $E_{\text{tot}}$ and the neutron
rms radius $r_{\mathrm{rms}}$ obtained by continuum and discretized HFB
calculations for $^{124}$Zr with varying DDDI pairing strengths $\eta V_{0}$, where $V_{0}=-458.4 $~MeV~fm$^{-3}$ and the factor $\eta=0.6\sim1.4$. For convenience, the corresponding neutron pairing energy ($-0.7\sim-83.7$~MeV)
and the average pairing gap defined as
\begin{equation}
\Delta_{uv}=\frac{\int\Delta(\mbox{\boldmath$r$})\tilde{\rho}(%
	\mbox{\boldmath$r$})~d\mbox{\boldmath$r$}} {\int\tilde{\rho}(%
	\mbox{\boldmath$r$})~d\mbox{\boldmath$r$}}
\end{equation}
($0.2\sim2.9$~MeV) are shown under the corresponding pairing strength factor
$\eta$.

One can see that, as the pairing strength increases, the total energy $E_{%
	\text{tot}}$ monotonously decreases: the nucleus becomes more bound, due to
the attractive pairing interaction. Moreover, one can find almost no
difference between the total energy obtained by the discretized and by the
full continuum HFB calculations. Naively thinking, the corresponding nuclear
size is shrinking as the nucleus becomes more bound. However, both the
discretized and the continuum calculations show that the neutron rms radius
first decreases, then reaches a local minimum at $\eta\approx 1.0$ and
afterwards increases. The difference between the discretized and continuum
results for the neutron rms radius is more obvious for weak pairing case
with a shallow Fermi energy.

\begin{figure}[ptb]
	\caption{Total neutron densities $4\protect\pi r^{2}\protect\rho(r)$ of $%
		^{124}$Zr obtained by the continuum HFB calculation with the SkI4
		interaction and the DDDI pairing force strengths $\protect\eta V_{0}$ for $%
		\protect\eta=0.6 \mathrm{~(dashed~line)},~1.0\mathrm{~(solid~line)},~1.4%
		\mathrm{~(dotted~line)}$. The insert shows the asymptotic behavior of the
		density in a logarithmic scale.}
	\label{fig2}
	\centering
	\includegraphics[width=0.5\textwidth]{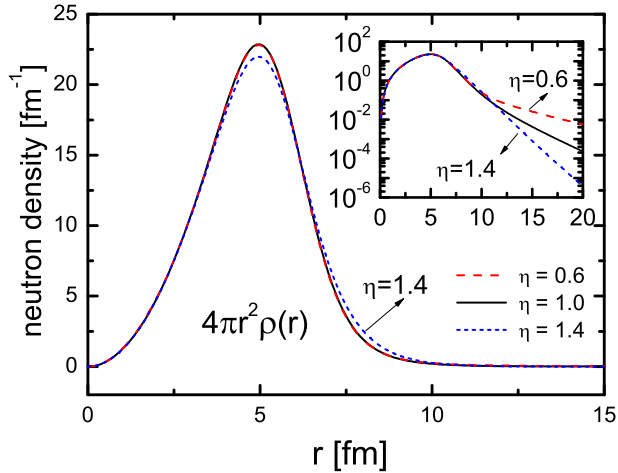}
\end{figure}

In order to understand the change of the rms radius as a function of the
pairing strength, we first plot in Fig.~\ref{fig2} the total neutron density
obtained by the continuum HFB calculation for the pairing strength factors $%
\eta =0.6$, $1.0$, and $1.4$. A linear scale is used in the main figure for
the inner and the surface region, and a logarithmic scale is used in the
insert for the asymptotic region.

In the insert we see that as the pairing strength increases, the density in
the asymptotic region ($r>10$~fm) always decays faster. This corresponds to
the `anti-halo effect' discussed in Ref.~\cite{Bennaceur2000_PLB496-154}
leading to a shrinkage of the rms radius. However, we also have to consider
the change of the density inside the nucleus and at the surface. From $%
\eta=0.6$ to $\eta=1.0$, the density inside the nucleus and at the surface
does not change much as shown in the main figure, but decays faster in the
asymptotic region as shown in the insert. As a result, the total rms radius
is decreasing up to $\eta=1.0$. When $\eta$ increases further, we observe a
change of the density in the surface region, first a small decrease around $%
r\sim 5$ fm and then a small increase in the region between $6 \lesssim r
\lesssim 10$ fm. For radii $r \gtrsim 10$ fm the density decreases again as
shown in the insert. However, the densities are so small at these large
radii, that this effect can only be recognized on the logarithmic scale and
it does not contribute much to the total radius. Therefore, the total radius
is determined by a competition between the increase of the density at the
surface and the decrease in the asymptotic region. Although the density
decays even faster in the asymptotic region, the increase at the surface $6
\lesssim r \lesssim 10$ fm finally produces an increase of the total rms
radius for $\eta > 1.0$.

\begin{figure}[ptb]
	\caption{{\protect\footnotesize (a) Single-neutron energies of $^{124}$Zr
			around the Fermi energy for $3p$, $2f$ and $1h$ orbits, which are the
			eigenvalues of the single-particle hamiltonian $h$ after the final
			convergence of the continuum Skyrme HFB calculation, and the dashed line
			denotes the Fermi energy $\protect\lambda$; (b) neutron occupation
			probabilities $v^2_{nlj}$, and (c) the contributions to the rms radius $r^{%
				\mathrm{rms}}_{nlj}$ from the quasiparticle states corresponding to the
			orbits shown in panel (a) within the energy $E=0\sim 6$~MeV in the continuum
			HFB calculation obtained with different pairing strength $\protect\eta V_{0}$%
			. }}
	\label{fig3}
	\centering
	\includegraphics[width=0.45\textwidth]{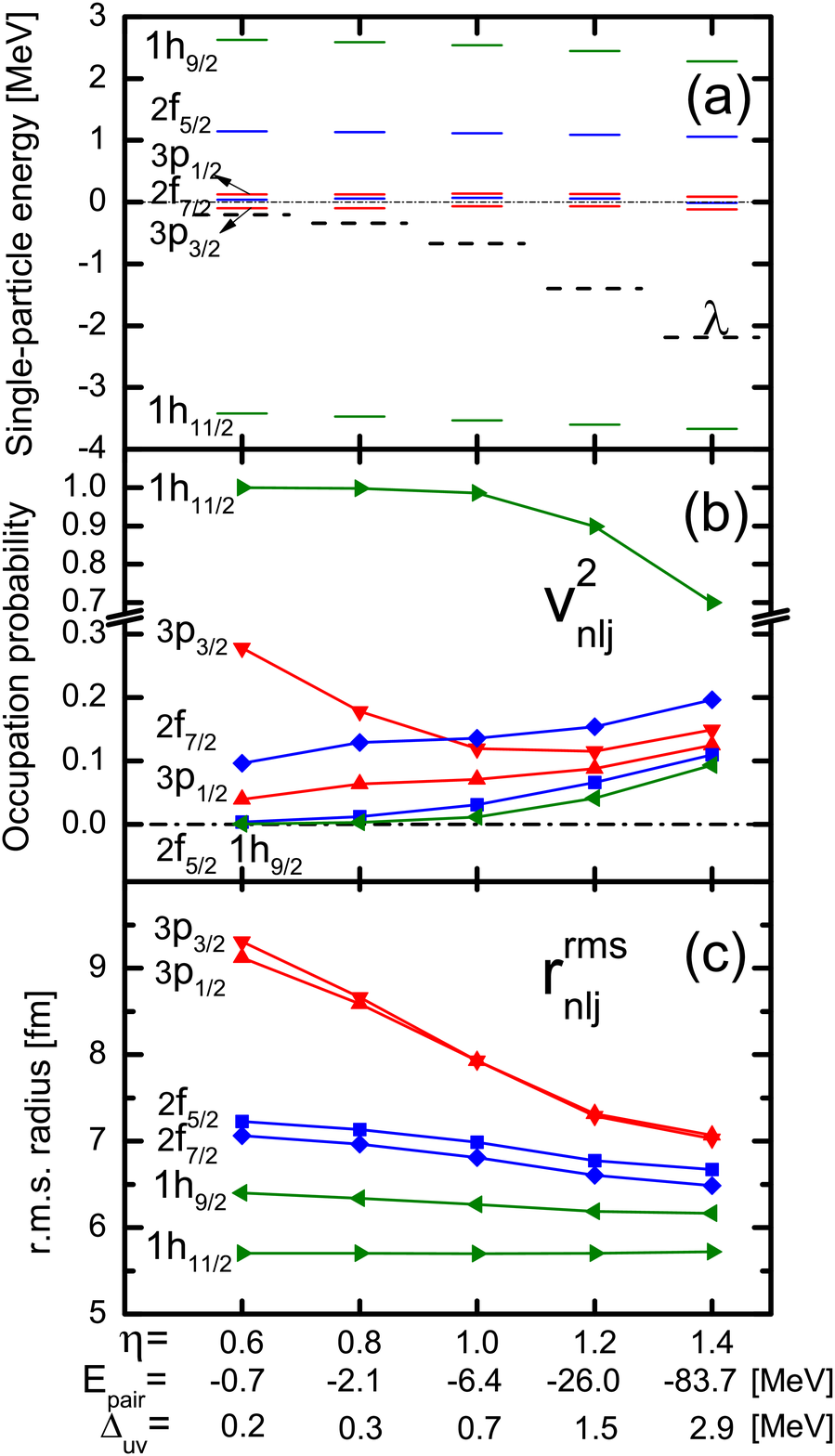}
\end{figure}

In order to understand the reason why there is an increase of the density at
the surface, we plot in Fig.~\ref{fig3} the information on important
single-neutron levels in the vicinity of the Fermi energy as function of
the pairing strength factor $\eta $.  In the discretized method the total
neutron rms radius of the nucleus is given as
\begin{equation}
\left( r_{\text{rms}}\right) ^{2}=\frac{\sum_{nlj}(2j+1)v_{nlj}^{2}\left(
	r_{nlj}^{\text{rms}}\right) ^{2}}{\sum_{nlj}(2j+1)v_{nlj}^{2}}.
\label{eq:rms-total}
\end{equation}%
It is determined not only by the rms radii $r_{nlj}^{\text{rms}}$ of the
individual orbits
\begin{equation}
r_{nlj}^{\mathrm{rms}}=\left( \frac{\int 4\pi r^{4}\rho _{nlj}(r)dr}{\int
	4\pi r^{2}\rho _{nlj}(r)dr}\right) ^{1/2},  \label{eq:rmsnlj}
\end{equation}%
but also by the occupation factors $(2j+1)v_{nlj}^{2}$ of these orbits
\begin{equation}
v_{nlj}^{2}=\frac{1}{2j+1}\int 4\pi r^{2}\rho _{nlj}(r)~dr.
\label{eq:occ-probability}
\end{equation}
Here the density distribution $\rho _{nlj}(r)$ of the orbit with the quantum
numbers ($nlj$) is given by the square of the corresponding quasiparticle wave function.

In our application the continuum is not discretized. We use the Green's
functions techniques. Here the sum over $n$ in Eq. (\ref{eq:rms-total}) is
replaced by a contour integration in the complex energy plane containing
all the bound states and the resonances.
The contributions of the individual orbitals to the density is calculated by the HFB Green's function constructed by the quasiparticle wave functions as
\begin{equation}
\rho _{nlj}(r)=\frac{1}{4\pi }\frac{1}{2\pi i}(2j+1)\oint_{C_{n}}\frac{%
	\mathcal{G}_{0,lj}^{(11)}(r,r,E)}{r^{2}}dE.  \label{eq:density-nlj}
\end{equation}
The integration is carried out on a
closed contour path $C_{n}$ \cite{Zhang-Ying2011_PRC83-054301},
choosing for each pair of quantum numbers $(lj)$ a rectangular path with
the energy interval $E=0\sim 6$~MeV. This path
includes for each $(lj)$-value the contribution of the lowest quasiparticle
state, i.e. the state closest to the Fermi energy, which is shown in Fig.
\ref{fig3}.

In Fig.~\ref{fig3} we give details on some important single-neutron levels around the Fermi energy in $^{124}$Zr as a function of the pairing strength factor $\eta$.  In principle, to investigate the single-particle properties, one should work in the canonical basis, where any HFB wave function can be expressed as a BCS wave function~\cite{Ring1980} and the occupation numbers $v_{nlj}^{2}$ are of BCS-form.  In this case $\varepsilon_{nlj} = <nlj|\hat{h}|nlj>$ is just the expectation value of the single-particle operator $\hat{h}$ in this basis. In the usual investigations working with a fixed box radius $R$~\cite{Meng1996_PRL77-3963, CHEN-Y2014_PRC89-014312} one has only discrete levels $\varepsilon_{nlj}$  and it is easy to find the canonical basis just by diagonalizing the density matrix $\rho^{lj}_{nn'}$.  In our method, since we construct only
the local density by the contour integral over the quasiparticle energy as shown in Eq.~(\ref{eq:density-nlj}),
we do not have access to the canonical basis for the moment. Instead, we show in Fig.~\ref{fig3}(a) the single-particle energies obtained by diagonalizing the single-particle Hamiltonian $\hat{h}$ on a mesh with a finite box size.  We use them as a reference to show the important levels around the Fermi energy in the present investigation~\cite{Meng1996_PRL77-3963, Dobaczewski1996_PRC53-2809}.  In the canonical basis, the occupation num¬bers $v^2_{nlj}$ depend in a sensitive way on the pairing correlations and the position of the corresponding energy levels $\varepsilon_{nlj}$.  But for the same reason, we instead use the neutron density $\rho_{nlj}(r)$ derived from the Green's function around the $n$th single-neutron state in Eq.~(\ref{eq:density-nlj}), to calculate the occupation numbers in Eq.~(\ref{eq:occ-probability}) shown in Fig.~\ref{fig3}(b).  They give a reference for the occupation situation of those important levels with different pairing strengths.  Figure~\ref{fig3}(c) shows the rms radius $r^{\rm rms}_{nlj}$ calculated by the same neutron den¬sity in Eq.~(\ref{eq:density-nlj}).

In the single-neutron spectrum in panel~(a), we find the weakly bound $%
3p_{3/2}$ orbit near the Fermi energy, the $2f_{7/2}$ and $3p_{1/2}$ orbits
just above the continuum threshold, and the $2f_{5/2}$ and $1h_{9/2}$ orbits
higher above. Without pairing, we know that the last two neutrons in $^{124}$%
Zr occupy the $3p_{3/2}$ orbit. With increasing pairing this level is more
and more depleted as shown in panel~(b). Only for $\eta > 1.2$ its
occupation probability slightly increases. At $\eta=0.6$ the Fermi energy is
located near this level with a rather small pairing energy ($-0.7$~MeV). As
the pairing strength increases, we find that the single-neutron energies
remain almost unchanged except for a little decrease for the $1h$ orbits. At
the same time, the pair scattering becomes stronger, and brings more
neutrons from the region below the Fermi energy to the region above it. As a
result, the occupation of the $1h_{11/2}$ level decreases and the Fermi
energy is pulled down closer to this orbit.
Panel~(b) shows the occupation probabilities $v_{nlj}^{2}$ of the $3p$, $2f$
and $1h$ orbits around the Fermi energy. Since the $s$ and $d$ orbits with
positive parity in the continuum have a much smaller occupation probability (%
$<2\%$), they are not shown in this figure. One can see that, as the pairing
strength increases, the neutrons in the $1h_{11/2}$ orbits are scattered up
to the $3p_{1/2}$, $2f_{7/2}$ orbits, and even to the $2f_{5/2}$ and $%
1h_{9/2}$ orbits high above in the continuum, while the neutron number for
the $3p_{3/2}$ orbit first decreases and then increases. Here, we should
notice that, the occupation probability calculated by Eq.~(\ref%
{eq:occ-probability}) counts not only the neutrons located on the
single-neutron levels, but also in the nearby continuum within the
quasiparticle energy $E=0\sim 6$~MeV, since these states become
quasiparticle resonant states with finite width due to the coupling with the
continuum by pairing~\cite{Zhang-Ying2011_PRC83-054301}.

The rms radii in Eq. (\ref{eq:rmsnlj}) for the corresponding orbits are shown in
panel (c). One can clearly see that, due to the lower centrifugal barrier,
the rms radii of the $3p$ orbits are larger than those of the $2f$ and $1h$
orbits, especially when the pairing strength is small. As the pairing
strength increases, due to the `pairing anti-halo effect', the rms radius
decreases for the $3p$ orbits dramatically, but it remains almost the same
for the $1h_{11/2}$ orbit. Thus at $\eta =1.4$ the rms radius of these
levels become comparable with each other. The smaller rms radius with a
stronger pairing gap seems to coincide with the so-called `pairing anti-halo
effect' shown by Eq.~(\ref{eq:rmsHFB}). However, we should keep in mind that
the total neutron rms radius (\ref{eq:rms-total}) is determined not only by
the rms radii $r_{nlj}^{\text{rms}}$ of the individual orbits in Eq.~(\ref%
{eq:rmsnlj}), but also by the occupation factors $(2j+1)v_{nlj}^{2}$ of
these orbits in Eq.~(\ref{eq:occ-probability}). Therefore it can happen, as
in the case of $^{124}$Zr, that all the individual rms radii, given in the
panel (c) of Fig.~\ref{fig3} decrease for $\eta >1.0$ and nonetheless the
total rms radius shown in Fig.~\ref{fig1} is increasing. The total neutron
density has contributions from all the levels weighted by the occupation
probabilities shown in panel (b). As the pairing strength increases, the
occupation probability of the $3p_{1/2}$, $2f_{7/2}$, and even $2f_{5/2}$ as
well as $1h_{9/2}$ orbits higher up in the continuum increases around $0.1$
respectively. This contributes to a larger total rms radius $r_{\mathrm{rms}}
$. As a result of the competition between the shrinkage of the rms radius
$r_{nlj}^{\mathrm{rms}}$ for the individual occupied single-particle orbits
and the growing occupation $v_{nlj}^2$ of low-$\ell$-orbits with large
rms radius in the continuum, the total rms radius first decreases and then increases as shown in Fig.~\ref%
{fig1}. Strong pairing correlations do not necessarily shrink the nuclear
size as it is indicated by the word `anti-halo'. It is essential to take
into account the change of the occupation among different orbits due to the
pair scattering, and especially the contributions from the continuum.

\begin{figure}[ptb]
	\caption{Neutron density $4\protect\pi r^{2}\protect\rho_{nlj}(r)$ of $3p$, $%
		2f$ and $1h$ orbits in $^{124}$Zr contributed from the quasiparticle states
		within the energy $E=0\sim 6$~MeV calculated by the continuum HFB approach
		with different pairing strengths $\protect\eta V_{0}$. The inserts give the
		same density distributions with logarithmic scale. In panel (f) for $%
		1h_{11/2}$ orbit, the neutron density is scaled by a factor $0.1$ compared
		to the original value. }
	\label{fig4}\centering
	\includegraphics[width=0.5\textwidth]{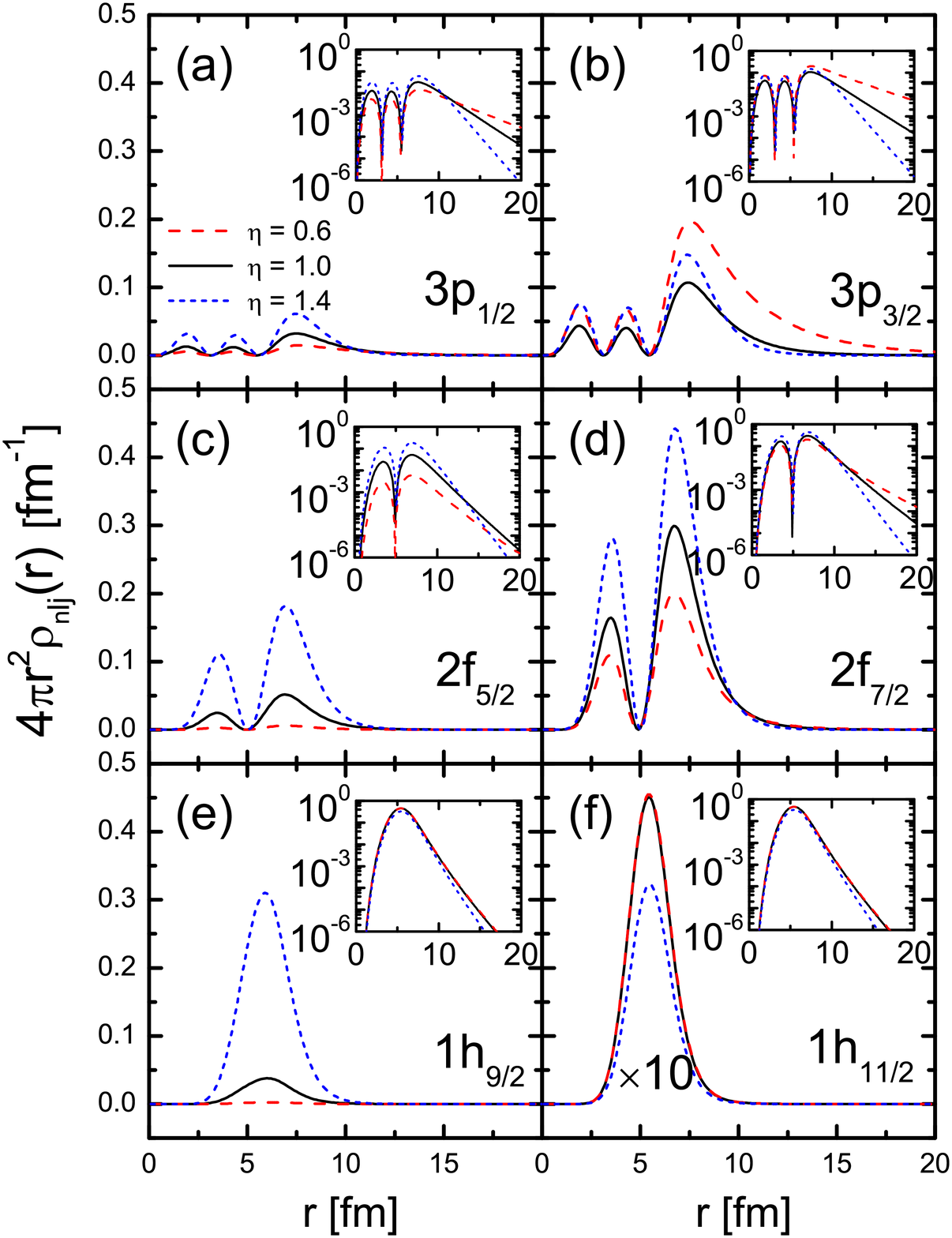}
\end{figure}

More explicitly, in Fig.~\ref{fig4} (a)-(f) we plot the neutron densities $%
4\pi r^{2}\rho_{nlj}(r)$ calculated by Eq.~(\ref{eq:density-nlj}) (including
the degeneracy factor $2j+1$) for $3p$, $2f$ and $1h$ orbits as
a function of the pairing strengths factor $\eta$. Again, the inner and
surface part of the densities are plotted in a linear scale in the main
figures, and the asymptotic behavior is plotted in a logarithmic scale in
the inserts. For comparison, we plot the densities for different orbits in
the same linear and logarithmic scales respectively, only the $1h_{11/2}$
orbit in panel (f) is scaled by a factor $0.1$.

Obviously, the $1h_{11/2}$ orbit contributes most to the density $\rho_{nlj}(r)$
within the quasiparticle energy interval $E=0\sim 6$~MeV shown in panel (f). It
remains almost the same from $\eta=0.6$ to $\eta=1.0$, but it drops
dramatically for $\eta=1.4$ at the peak near $r\approx 5$~fm, which leads
to the reduction of the occupation probability shown in Fig.~\ref{fig3} (b).
Actually, this drop of the density at the peak is also the main reason for
the decrease of the total density around $r\approx 5 $~fm shown in
Fig.~\ref{fig2}.

The contribution to the density $\rho_{nlj}(r)$ from the $3p_{3/2}$ orbit in
panel (b) decreases obviously at the surface between $\eta=0.6$ and
$\eta=1.0 $, which leads to the reduction of its occupation probability shown
in Fig.~\ref{fig3} (b). With further stronger pairing at $\eta=1.4$, the
density at the surface increases again due to the neutrons
scattered from below (e.g., from the $1h_{11/2}$ orbit). In the asymptotic region,
the density contribution always decays faster with increasing pairing
strength for all values of the parameter $0.6\leq\eta\leq 1.4$. Moreover, one should notice that it
is the $3p_{3/2} $ state that dominates the density in the asymptotic region
and thus governs the asymptotic behavior of the total density for $r\gtrsim10$%
~fm shown in the insert of Fig.~\ref{fig2}.

Another large density contribution comes from the $2f_{7/2}$ orbit in panel
(d). It has a dramatic increase inside and around the surface, with stronger
pairing strength. This helps to cancel the effect of the decreasing contribution from
the $3p_{3/2}$ orbit.

For the $3p_{1/2}$, $2f_{5/2}$, and $1h_{9/2}$ orbits shown in panels (a),
(c), and (e), one finds an obvious increase from almost zero for the density
contribution especially around the surface. This increase indicates that
neutrons begin to occupy these orbits when the pairing scattering is strong
enough. Together with the contribution from the $2f_{7/2}$ orbit, we can explain
the increase of the total density at the surface ($r=6\sim8$~fm) for $\eta=1.4$
shown in Fig.~\ref{fig2}, which at last leads to the increase of the total
rms radius at $\eta=1.4$ shown in Fig.~\ref{fig1}.

So far, taking the nucleus $^{124}$Zr as an example, we change the pairing
strength by an arbitrary factor $\eta=0.6\sim1.4$, and see what happens to
the neutron rms radius. Of course, the arbitrary change of the pairing
strength is not physical, but it can serve as a model analysis of this
problem. In order not to deviate from the physical point too much, we keep
ourself not so far away from $\eta=1.0$. Actually, in Ref.~\cite%
{Bennaceur2000_PLB496-154}, where the `pairing anti-halo effect' is
discussed, the rms radius derived from HF wave functions without pairing is
compared with that derived from HFB wave functions with pairing. This is not
exactly what we have done above. In order to compare with the no-pairing
case, we also performed a HF calculation for this nucleus without pairing
($\eta=0$), and obtain the neutron rms radius $r^{\mathrm{HF}}_{\mathrm{rms}%
}=5.08$~fm, which is even $0.6\%$ larger than the continuum HFB result
$r^{\mathrm{HFB}}_{\mathrm{rms}}=5.05$~fm at $\eta=0.6$. From the exact
no-pairing (HF) case to the small pairing (HFB) case ($\eta<1.0$), the
decrease of the neutron rms radius is mainly due to the shrinkage of the rms
radius of the occupied single-particle orbits ($3p_{3/2}$), which seems to
coincide with the so-called `pairing anti-halo effect'.

\begin{figure}[ptb]
	\caption{(a)~Total energy $E_{\text{tot}}$ and the neutron rms radius $r_{%
			\mathrm{rms}}$ obtained by the continuum (filled circle) and the discretized
		(open circle) HFB calculations for $^{122}$Zr, and (b) occupation
		probabilities of the $3p$, $2f$ and $1h$ orbits around the Fermi energy
		contributed from the quasiparticle states within $E=0\sim 7$~MeV obtained by
		the continuum HFB calculation with the SkI4 interaction and the DDDI pairing
		force with different strengths $\protect\eta V_{0}$.}
	\label{fig5}
	\centering
	\includegraphics[width=0.5\textwidth]{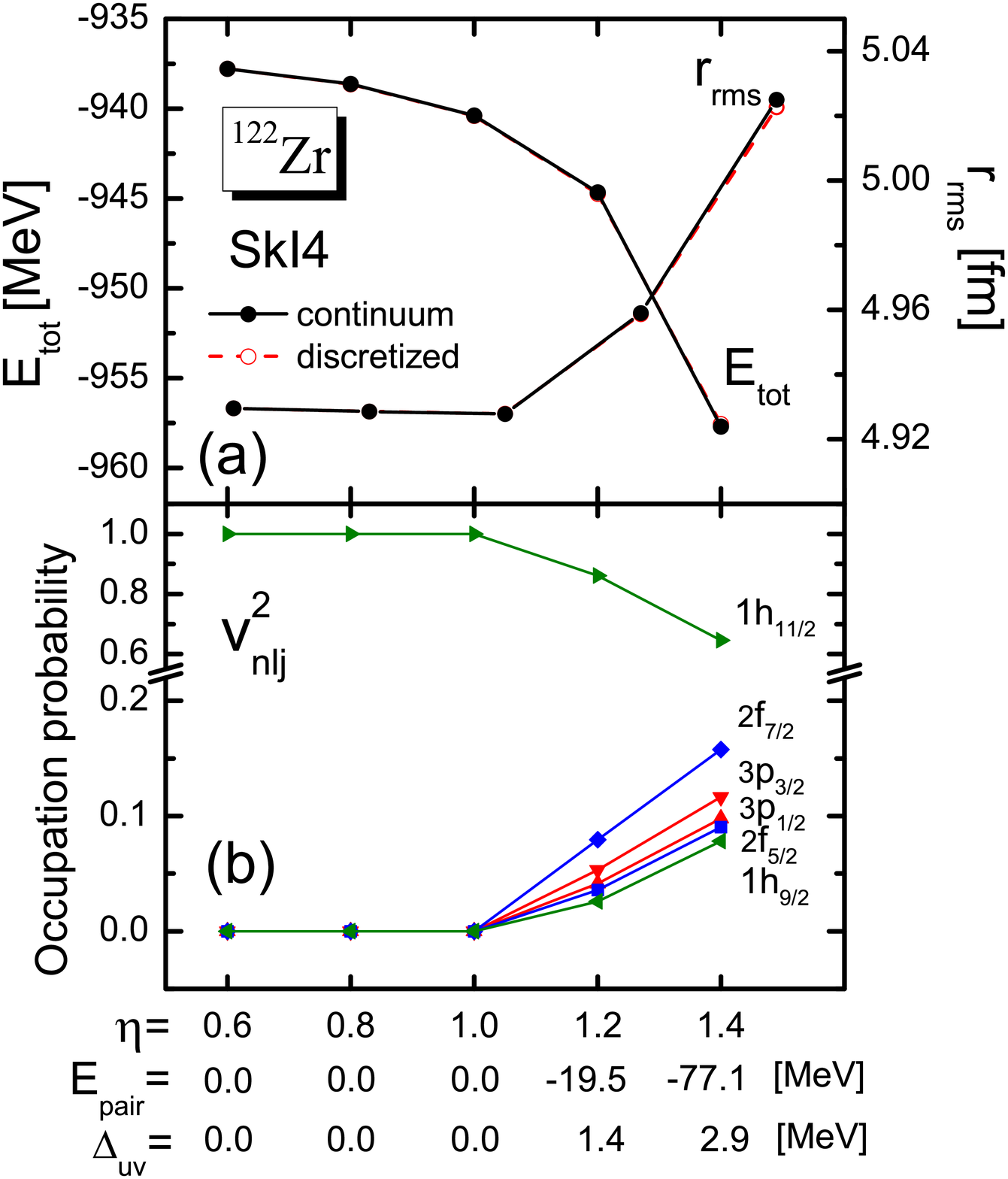}
\end{figure}

However, if we examine another example, $^{122}$Zr, we can not find such a
`pairing anti-halo effect' at all.
Here, we do not claim $^{122}$Zr is a real halo nucleus.  We will check the change of the rms radius with the increasing pairing strength.
The total energy and the neutron rms
radius are shown in Fig.~\ref{fig5} (a) as a function of the pairing strength
factor $\eta$. It is clearly seen that as the nucleus becomes more bound
with stronger pairing correlations, the neutron rms radius first remains unchanged
and then monotonously increases. In panel (b) we show the corresponding occupation
probabilities of the $3p$, $2f$, and $1h$ orbits,  coming from the quasiparticle
states within $0\leq E\leq 7$~MeV as a function of the pairing strength factor $\eta$.
It is clear that, without pairing, the $1h_{11/2}$ orbit is fully occupied.
For a small pairing strength, the energy gap around $3$~MeV (see Fig. \ref{fig3} (a)) makes
it difficult to scatter neutrons to higher orbits. Therefore the HFB result
for the rms radius remains almost the same for all $\eta$-values
$0\leq\eta\leq1.0$, similar to the HF case with a vanishing pairing energy. For a
further increase of the pairing strength, the rms radius of the occupied $%
1h_{11/2}$ orbit remains almost the same (see Fig.~\ref{fig3} (c)). However,
the neutrons begin to occupy the $3p$ and $2f$ orbits, which can contribute
a larger rms radius $r^{\mathrm{rms}}_{nlj}$ as shown in Fig.~\ref{fig3}
(c). Therefore in this case, we can only observe an expansion but no
shrinkage of the neutron rms radius from the no-pairing to the finite
pairing case. Actually, the difference between $^{124}$Zr and $^{122}$Zr is
whether the weakly bound $3p_{3/2}$ orbit is originally occupied or not. The asymptotic
behavior of this wave function is sensitive to the pairing gap.

Finally, we would like to emphasize that the terminology
`pairing anti-halo effect' has to be taken with great care. In the original
literatures~\cite{Bennaceur2000_PLB496-154}, the `pairing anti-halo effect'
is actually referred to the fact that the divergence of the HF wave function
without pairing at the limit $\varepsilon\to0$ can be avoided in the HFB
wave function with finite pairing gap. It is restricted to the rather
academic case of one fully occupied single-particle wave functions with low-$%
\ell$ value very close to the continuum limit. The realistic case is much
more complicated. In particular one has to consider contributions of several
partially occupied orbits and the occupation are determined by the pairing
correlations. This can lead to a decrease or an increase of the various
contributions. In some cases it is just the influence of pairing, which causes
an extension of the neutron radius in such nuclei.

\section{Conclusions}
In the present investigations, we concentrated on the influence of pairing
correlations on the size of the neutron-rich nuclei using a fully self-consistent
description as well for the mean potential as for the pairing correlations.
Taking the nucleus $^{124}$Zr as an example, we performed a numerical analysis
by self-consistent Skyrme HFB calculations varying the strength of pairing
in reasonable limits. The continuum is treated appropriately by Green's
function techniques. We find that the neutron rms radius first shrinks and then
expands as the pairing strength increases. The expansion of the neutron rms radius by
pairing seems to contradict to the so-called `pairing anti-halo effect',
which is associated with a reduction of the halo size by the pairing
correlations.

In fact, it is clear that stronger pairing causes the density of specific
orbits to decay faster in the asymptotic region, especially for occupied low
angular momentum states. This can lead to a shrinkage of the total rms
radius. However, at the same time, one should also take into account the
reoccupation processes caused by pairing and the corresponding coupling to
the continuum. There exist `halo' orbits, i.e. weakly bound states and
resonant states embedded in the continuum which, by themselves, have a large
rms radius. Without pairing they are unoccupied and cannot contribute to the
total radius. On the other hand the self-consistent solution of the
corresponding HFB equations in the continuum leads to changes in the
occupation pattern. This will
lead to an increasing of the radius, which would not exist without pairing.

In the self-consistent calculation of our example nucleus $^{124}$Zr, the
above two aspects compete with each other and cause with increasing pairing
strength the total neutron rms radius first to decrease and then to
increase. The terminology `pairing anti-halo effect' only emphasizes the
first aspect. Therefore, this terminology is somehow misleading for this
purpose. After all, `halo' is not equivalent to `divergence', and pairing
correlations play an important role in the anomalous increase of neutron radii
in such nuclei.

\begin{acknowledgments}
The author Y.Z. would like to thank L.L. Li and S.G. Zhou for helpful
discussions. This work was partly supported by the Major State 973 Program
2007CB815000; the National Natural Science Foundation of China under No.
11405116, No. 11335002, and No. 11175002; the Research Fund for the Doctoral
Program of Higher Education No. 20110001110087; and the Oversea
Distinguished Professor Project from Ministry of Education No.
MS2010BJDX001. We also acknowledge partial support from the DFG Cluster of
Excellence Origin and Structure of the Universe (www.universe-cluster.de).

\end{acknowledgments}

\end{document}